\documentclass{elsart}

\usepackage{graphicx}
\usepackage{amsmath,amsfonts,amssymb}
\usepackage[english,francais]{babel}
\newtheorem{e-proposition}[theorem]{Proposition}

\newtheorem{e-definition}[theorem]{Definition\rm}

\newcommand{\simgt}{\stackrel{>}{{}_\sim}}

\def\og{\leavevmode\raise.3ex\hbox{$\scriptscriptstyle\langle\!\langle$~}}
\def\fg{\leavevmode\raise.3ex\hbox{~$\!\scriptscriptstyle\,\rangle\!\rangle$}}

\begin{document}
% Select a primary header Physics or Astrophysics
% You can place after the header (classification), if you know it.

\centerline{Physics/Header}
\begin{frontmatter}


% \author{Name\thanksref{label2}}
\selectlanguage{english}
\title{Universal few-body physics in a harmonic trap}

% PREPRINT HISKP-TH-10-20

% If all authors are at the same address, the [label1] can be suppressed

\selectlanguage{english}
\author[authorlabel1]{S. T\"olle},
\ead{toelle@hiskp.uni-bonn.de}
\author[authorlabel1]{H.-W. Hammer},
\ead{hammer@hiskp.uni-bonn.de}
\author[authorlabel1]{B.Ch. Metsch}
\ead{metsch@hiskp.uni-bonn.de}

\address[authorlabel1]{
Helmholtz-Institut f\"ur Strahlen- und Kernphysik
        and Bethe Center for Theoretical Physics,
        Universit\"at Bonn, 53115 Bonn, Germany
}
%\address[authorlabel2]{Address2}

% If your know the dates of reception, and acceptation you can put them now;
%    idem the name of the person presenting your article

\medskip
\begin{center}
{\small Received *****; accepted after revision +++++}
\end{center}

\begin{abstract}
Few-body systems with resonant short-range interactions display 
universal properties that do not depend on the details of their 
structure or their interactions at short distances. 
In the three-body system, these properties include the existence 
of a geometric spectrum of three-body Efimov states and a discrete 
scaling symmetry. Similar universal properties appear in 4-body and 
possibly higher-body systems as well. 
We set up an effective theory for few-body systems in a harmonic
trap and study the modification of universal physics for 3- and 
4-particle systems in external confinement. In particular, we focus on
systems where the Efimov effect can occur and
investigate the dependence of the 4-body spectrum on the 
experimental tuning parameters.
%{\it To cite this article: A. Name1, A. Name2, C. R. Physique 6 (2005).}

\vskip 0.5\baselineskip

\selectlanguage{francais}
\noindent{\bf R\'esum\'e}
\vskip 0.5\baselineskip
\noindent
{\bf Physique universelle de quelques corps dans un pi\`ege harmonique}

\noindent
Des syst\`emes \`a quelques corps avec des interactions resonantes aux courtes distances 
montrent des propri\'et\'es universelles qui ne dependent pas des d\'etails de
leur structure ou de leurs interactions  aux courtes distances. Dans des syst\`emes
\`a trois corps, cettes propri\'et\'es incluent l'existence d'un spectre
g\'eom\'etrique d'\'etats d'Efimov \`a trois corps et une sym\'etrie d'echelle
discr\`ete.
Des propri\'et\'es universelles similaires se trouvent dans des syst\`emes \`a quatre
corps et possiblement dans des syst\`emes \`a davantage corps aussi. 
Nous construisons une th\'eorie effective pour des syst\`emes \`a quelques
corps dans un pi\`ege harmonique et \'etudions 
la modification de la physique universelle des syst\`emes \`a trois et
quatre corps dans un confinement ext\'erieur.  En particulier, nous nous concentrons 
sur les syst\`emes o\`u l'effet d'Efimov peut appara\^itre et nous examinons 
la d\'ependance du spectre \`a quatre corps aux
param\`etres experimentels ajustables. 
%{\it Pour citer cet article~: A. Name1, A. Name2, C. R. Physique 6 (2005).}

%Now keywords/mots-clÈs
 \vskip 0.5\baselineskip
\noindent{\small{\it Mots-cl\'es~:} Universalit\'e~; Th\'eorie Effective~;
Confinement Ext\'erieur}}
\end{abstract}
\end{frontmatter}

% now the Version franÁaise abrÈgÈe, if it exists
%\selectlanguage{francais}
%\section*{Version fran\c{c}aise abr\'eg\'ee}
% Text of your Version franÁaise abrÈgÈe here

\selectlanguage{english}
% main text
\section{Introduction}
\label{sec:intro}
% etc, etc
Few-body systems close to the unitary limit show interesting 
universal properties. In the 2-body system, the scattering amplitude
saturates the unitarity bound from conservation of probability.
Such systems  are characterized by a scattering length $a$ that
is much larger than the typical range of the interaction $r_0$
 and $a$ is the only relevant length scale at low energies.
If $a$ is positive, 
two particles of mass $m$ form a shallow dimer with energy
%\beq
$E_2 \approx -{\hbar^2}/{(m a^2)}\,$,
%\eeq
independent of the mechanism responsible for the large scattering length.
Examples for such shallow dimer states are the deuteron 
in nuclear physics, the $^4$He dimer in atomic physics, 
cold atoms close to a Feshbach resonance,
and  possibly the new charmonium state $X(3872)$ in particle 
physics \cite{Braaten:2004rn,Hammer:2010kp}.  

In the 3-body system, an additional length scale $1/\kappa_*$ 
is generated by the Efimov effect \cite{Efimov-70}.  
If at least two of the three pairs of particles have a large scattering 
length $|a| \gg r_0$, the Efimov effect can occur.
In the limit $1/a \to 0$, there
are infinitely many 3-body bound states with an accumulation point at
the 3-body scattering threshold. These Efimov states or trimers have a
geometric spectrum \cite{Efimov-70}:
%----------------------
\begin{eqnarray}
E^{(n)}_3 = -(e^{-2\pi/s_0})^{n-n_*} \hbar^2 \kappa^2_* /m,
\label{kappa-star}
\end{eqnarray}
%----------------------
which is specified by the binding momentum $\kappa_*$ of the Efimov trimer
labeled by $n_*$. This spectrum is a consequence of a discrete scaling
symmetry with discrete scaling factor $e^{\pi/s_0}$.  In the case of
identical bosons, $s_0 \approx 1.00624$ and the discrete scaling
factor is $e^{\pi/s_0} \approx 22.7$.  
The scaling symmetry becomes 
also manifest in the log-periodic dependence of 
scattering observables on the scattering length $a$ \cite{Efimov79}.  The
consequences of discrete scale invariance and \lq\lq Efimov physics'' can be
calculated in an effective field theory for short-range interactions,
where the Efimov effect appears as a consequence of a renormalization
group limit cycle \cite{Bedaque:1998kg}.

While the Efimov effect was established theoretically already in 1970, the
first experimental evidence for an Efimov trimer in a trapped gas of 
ultracold Cs atoms was
provided only recently by its signature in the 3-body recombination rate
\cite{Kraemer-06}. It could be unravelled by varying the
scattering length $a$ over several orders of magnitude using a
Feshbach resonance
and testing the predictions for the line shape of the loss resonance.
Since this pioneering experiment, there was 
significant experimental progress in observing Efimov 
physics in ultracold quantum gases.
More recently, evidence for Efimov trimers in 3-body recombination 
was also obtained in a balanced mixture of atoms in three
different hyperfine states of $^6$Li \cite{Ottenstein08,Huckans08},
in a mixture of Potassium and Rubidium atoms \cite{Barontini09},
and in an ultracold gas of $^7$Li atoms \cite{Gross09}. In another experiment
with Potassium atoms \cite{Zaccanti09}, two bound trimers were observed, 
whose energies are consistent with the predicted scaling relation.
Efimov states can also be observed as resonances in atom-dimer scattering. 
Such resonances have been seen using atom-dimer mixtures of Cs atoms
\cite{Knoop08} and of  $^6$Li atoms \cite{Lompe:2010,Nakajima:2010}.
The first direct observation of Efimov trimers of  $^6$Li atoms created by 
radio frequency association was recently reported by the Heidelberg 
group \cite{Lompe:RF}.

One of the most exciting recent developments in 
universal few-body physics involves
universal tetramer states. There is a pair of universal tetramer states 
associated with every Efimov trimer \cite{Hammer:2006ct,Stecher:2008}. 
The tetramer states above the ground state trimer acquire a width from the possible decay into a 
trimer and an atom. Deltuva has calculated these widths and found them to be small and
universal: They are 0.3\% of the binding energy for the deeper
state and 0.02\% of the binding energy for the shallower state \cite{Deltuva:2010xd}.
The resonant 
enhancement of 4-body recombination provides a signature for these tetramers
similar to the trimers \cite{Stecher:2008}.  
Loss resonances from both tetramers were subsequently 
observed in an ultracold gas of 
$^{133}$Cs atoms \cite{Ferlaino:2009}.  In $^7$Li atoms, even two sets of 
tetramers that are close to the corresponding Efimov trimers could be
observed \cite{Pollack:2009}.
Moreover, there is some theoretical evidence of even higher-body 
universal states \cite{Stecher:2009}.

These experiments were carried out in a regime where the 
influence of the trap on the few-body spectra could be neglected. 
However, the trap also offers new possibilities to modify 
the properties of few-body 
systems. In particular, a narrow harmonic confinement
changes the spectrum and can lead to interesting new phenomena. 
The 2-body problem in an isotropic harmonic trap was solved analytically by Busch
et al. \cite{busch}. 
The corresponding 3-body problem for spinless bosons was first solved by Jonsell et al.
\cite{jonsell}, while the case of  two-component fermions was considered by Tan \cite{tan}.
In the unitary limit of infinite scattering length, Werner and Castin 
calculated the complete 3-body spectrum for two-component fermions and bosons 
and provided an analytic solution \cite{werner}.
There are also a number of numerical studies of few-body systems with three and more particles
in harmonic confinement. Most of this work, however, has focused on the problem of two 
component fermions in or near the unitary limit 
\cite{chang,stecher:2007,blume:2007,Stetcu:2007ms,kestner:2007,Luu:2006xv,stecher:2008,Alhassid:2008,Stetcu:2010xq,Rotureau:2010uz,Drummond:2010}.

In this work, we investigate universal few-body physics in 
harmonic confinement for three and four particles. 
Our strategy follows Stetcu et al.~\cite{Stetcu:2006ey}, 
where an effective theory for
short-range forces in the framework of the no-core shell model was formulated.
The effective interactions were defined within a finite model space
with a cutoff $N$ on the basis functions. 
This strategy was also applied
to atomic systems of three and four spin-1/2 fermions in a trap
 \cite{Stetcu:2007ms}. 
In \cite{Stetcu:2010xq,Rotureau:2010uz}, they found that using
different cutoffs for systems with different  number of 
particles leads to improved 
convergence of perturbative higher-order corrections. This is because the 
spectator particles in a many-body system can carry some of the excitation
energy, leaving less available to an interacting pair. 
For sufficiently large cutoffs, however, the convergence problems disappear.
Here, we stay at leading order in the effective theory
and calculate the spectra of 
2-, 3- and 4-particle systems in a harmonic trap. We focus on systems
displaying the Efimov effect and perform a systematic 
study of universal bound state properties. In particular,
we investigate how the 
bound state spectrum depends on the 2- and 3-body energies required as input. 
Some preliminary results of our study were already reported in 
\cite{fb19toelle}.

The paper is organized as follows. In the next section, we give 
a brief overview of our 
effective theory for few-body systems with large scattering lengths in an
external confinement. In Sec.~\ref{sec:3body}, we present our results 
for 3-body systems in the unitary limit and compare with the analytical 
solution of Werner and Castin \cite{werner}. Moreover, we calculate the 
energies of the first two $L^P=0^+$ states for a 3-body system of $^6$Li
in the lowest two spin states. In Sec.~\ref{sec:4body}, we present 
our results for 3- and 4-body systems away from the unitary limit in
the form of an extended Efimov plot. We consider systems of
two identical fermions with two other distinguishable particles as well
as systems of identical bosons. The paper then ends with a 
summary and conclusions.

\section{Effective theories for trapped systems}
\label{sec:et}
In this section, we formulate our effective theory for particles with large $S$-wave scattering length 
in a harmonic trap. 
Following Ref.~\cite{Stetcu:2006ey}, we identify the ultraviolet cutoff of the effective theory with the 
model space cutoff on the harmonic oscillator basis functions.
Since the trap only modifies the infrared properties of the system, the renormalization in the ultraviolet 
remains the same and we require both a 2- and a 3-body interaction at leading order in the large scattering
length  \cite{Bedaque:1998kg}. 
The leading corrections are due to the range $R$ of the underlying interaction.
For typical momenta $\kappa\sim\sqrt{m|E|}$ of order $1/a$, these corrections are suppressed by
$R/a$. For broad enough Feshbach resonances, one can take $R$ to be equal to the van der Waals length $l_{vdW}$.
In the trap there are also corrections of order $R/\beta$, where
$\beta$ is the oscillator length defined below.
If large energies are involved, there are also corrections of order $R\sqrt{m|E|}$. 
The explicit consideration of these corrections will be left for a 
future publication.
For a discussion of higher order corrections for fermions with two spin 
states, see Refs.~\cite{Stetcu:2010xq,Rotureau:2010uz}.

For $A$ bodies with equal masses in an isotropic harmonic oscillator potential (HOP), 
the energy spectrum is determined by the Hamiltonian
\begin{align}
H=\sum_{i=1}^{A}\left(\frac{\bigl|\vec{p}_i\bigr|^2}{2m}+\frac{1}{2}m\omega^2\bigl|\vec{x}_i\bigr|^2\right)
+\sum_{i<j}^AV_{ij}+\sum_{i<j<k}^AW_{ijk}\:,
\label{eq:effham}
\end{align}
where $\omega$ is the trapping frequency and $V_{ij}$ and $W_{ijk}$ are 2- and 3-particle contact interactions between 
bodies $i$,$j$, and $k$. Since the interactions only depend on relative coordinates, Jacobi coordinates $\vec{s}_i$ 
are introduced and the dynamics of the centre of mass separate. As in free space, the ultraviolet
behavior of this Hamiltonian has to be regularised. 
Thus the Hilbert space is restricted by delimiting the basis functions with a regulator $N$. For $A$ bodies in the HOP 
it is convenient to use the tensor product of harmonic oscillator functions (HOF) $\phi_{n_il_im_i}\left(\vec{s}_i\right)$ 
as a basis. Then the model space for a given $N$ is the linear hull of HOF with the requirement 
$\sum_{i=1}^{A-1}\left(2n_i+l_i\right)\leq N$ \cite{Stetcu:2006ey}.
This means, the basis consists of states with an unperturbed eigenenergy less or equal to
$E_N=\sum_{i=1}^{A-1}(2n_i+l_i+3/2)\hbar\omega$.
In this sense, the regulator $N$ corresponds to 
an energy cutoff. We identify this cutoff with the ultraviolet cutoff of our effective theory. Since the model space 
is finite, the Hamilton matrix is finite, too. Thus the energy eigenvalues in 
the model space can be calculated by a numerical diagonalisation of the Hamilton matrix. Note that in free space one usually uses 
a momentum cutoff $\Lambda$ to regularize the resulting 
integral equations~\cite{Braaten:2004rn,Bedaque:1998kg}. For large values of $N$, $\Lambda$ is
proportional to $\sqrt{N}$.
We note that the 3-body spectrum is bounded from below even in the presence of the Efimov effect because
of the finite cutoff $N$. If $N$ was increased, lower and lower energy states would appear, but they 
would be outside of the range of applicability of our effective theory which cannot describe states
with $\sqrt{m|E|} \simgt 1/R$, where $R$ is the range of the underlying interaction.

\subsection{Two bodies}

In the 2-body sector the elements of the Hamilton matrix are
\begin{align}
\bigl<nlm\bigl|H^{\left(2\right)}\bigr|n'l'm'\bigr>=
\hbar\omega\left(2n+l+\frac{3}{2}\right)\delta_{n,n'}\delta_{l,l'}\delta_{m,m'}+\frac{v}{\pi^\frac{3}{2}}f_nf_{n'}
\delta_{l,0}\delta_{l',0}\:
\end{align}
with $f_n=\sqrt{\frac{\left(2n+1\right)!!}{n!2^n}}$. The running coupling constant $v(N)$ is renormalised by the 
requirement that the 
ground state energy in the model reproduces a given value. The separability of the interaction allows one to find the 
relation between the ground state energy $E_0^{(2)}=\epsilon^{\left(2\right)}\:\hbar\omega$, the cutoff parameter $N$ and 
the running coupling constant $v(N)$:
\begin{equation}
\label{2renorm}
-\frac{\pi^\frac{3}{2}\hbar\omega}{v}=\sum_{n=0}^{\frac{N}{2}}\frac{\left|f_n\right|^2}{2n+\frac{3}{2}
-\epsilon^{\left(2\right)}}\:.
\end{equation}

%%%%%%%%%%%%%%%%%%%%%%%%%%%%%%%%%%%%%%%%%%%%%%%%%%%%%%%%%%%%%%%%%%%%%%%%%%%%%%%%%%%%
\begin{figure}[t]
	\centering
	\includegraphics[width=0.5\linewidth, angle=270]{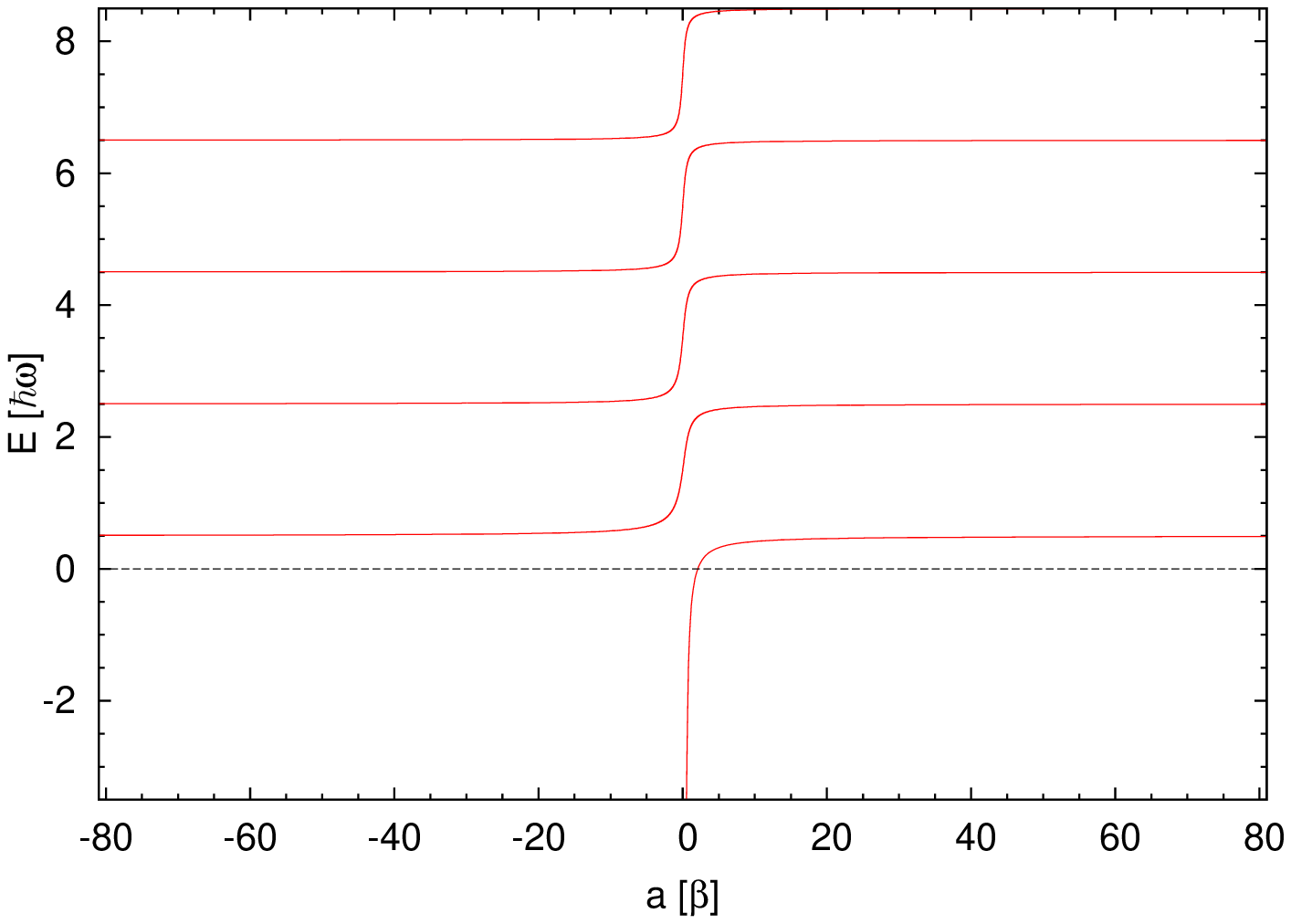}
	\caption{Energy spectrum for two particles in a harmonic trap as a function of the
           scattering length $a$ in units of the oscillator parameter $\beta$ according to \cite{busch}.}
	\label{fig:2ab}
\end{figure}
%%%%%%%%%%%%%%%%%%%%%%%%%%%%%%%%%%%%%%%%%%%%%%%%%%%%%%%%%%%%%%%%%%%%%%%%%%%%%%%%%%%%
The energy spectrum of the 2-body system in an isotropic harmonic trap is known exactly \cite{busch}. 
In Fig.~\ref{fig:2ab}, we show the energy spectrum as a function of the scattering length $a$ 
in free space measured in units
of the oscillator length $\beta=\sqrt{\hbar/(m\omega)}$.
In the unitary limit, the ground state energy approaches $\hbar\omega/2$.

%%%%%%%%%%%%%%%%%%%%%%%%%%%%%%%%%%%%%%%%%%%%%%%%%%%%%%%%%%%%%%%%%%%%%%%%%%%%%%%%%%%%
\begin{figure}[t]
	\centering
	\includegraphics[width=0.5\linewidth, angle=270]{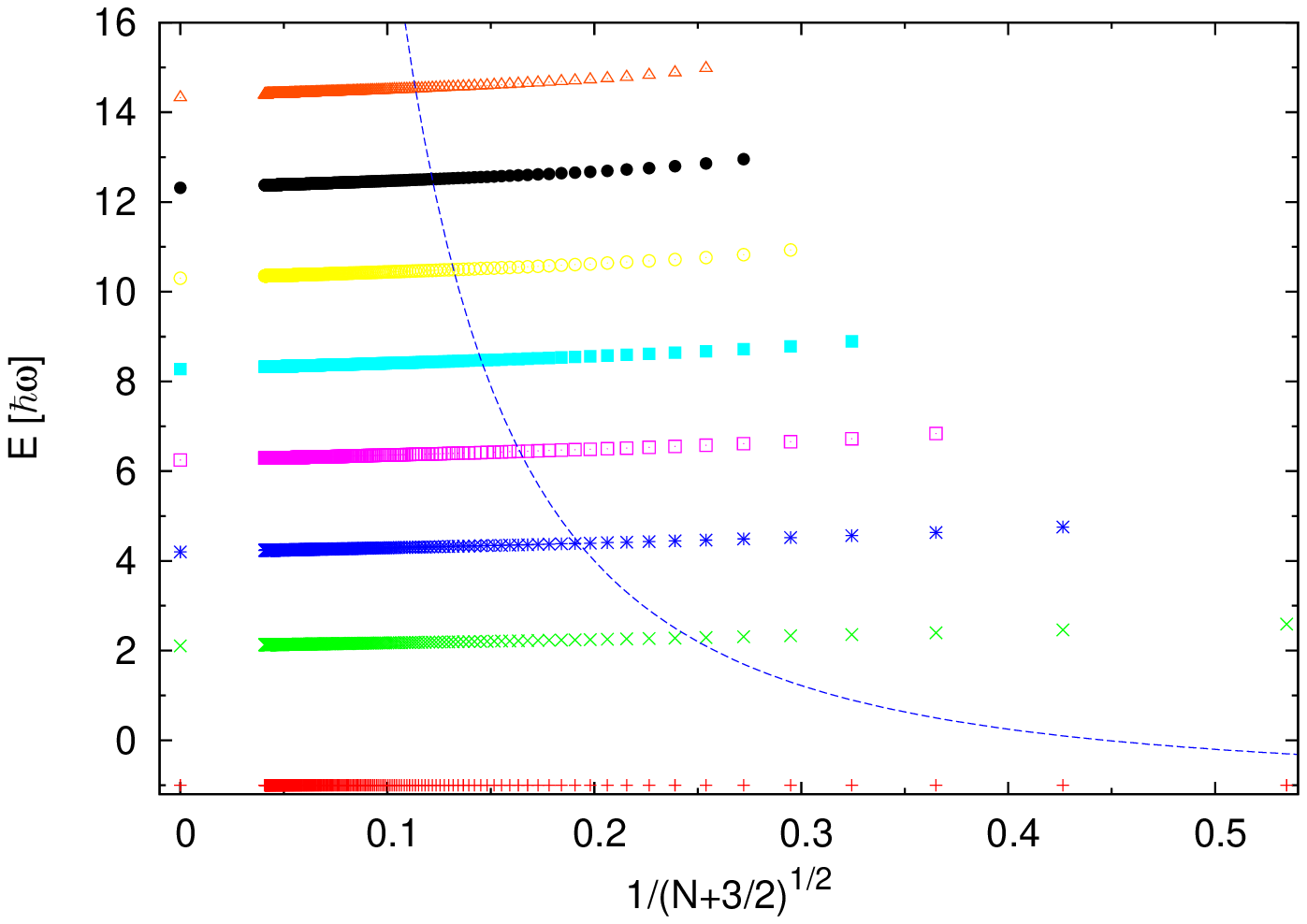}
	\caption{Energy spectrum for two particles in a harmonic trap as a function of the cutoff 
$N$ and exact results from \cite{busch} at $N\rightarrow\infty$ for $E_0=-\hbar \omega$.
       The dashed line indicates the value of the cutoff $N$ at which $N\approx 5 E/(\hbar\omega)$.}
	\label{fig:2spectrum}
\end{figure}
%%%%%%%%%%%%%%%%%%%%%%%%%%%%%%%%%%%%%%%%%%%%%%%%%%%%%%%%%%%%%%%%%%%%%%%%%%%%%%%%%%%%
We compare our numerical results to the exact solution in order to test the accuracy of
our method. In Fig.~\ref{fig:2spectrum}, we show
the results for the calculated spectra for $\epsilon^{\left(2\right)}=-1$ and cutoff parameters $N\leq600$. 
The exact results \cite{busch} are added at $N\rightarrow\infty$.
Already for finite values of the cutoff $N$ with $N\hbar\omega$ large compared to a calculated energy, we observe
good agreement with the exact solution. The exact values can be reproduced
numerically by extrapolating the results in the model spaces. We note, however, that this extrapolation is not
strictly necessary if we interpret Eq.~(\ref{eq:effham}) as an effective Hamiltonian that is only accurate up
to errors of order $\sqrt{E/(N\hbar\omega)}$. 
The dashed line indicates the value of the cutoff $N$ at which
$N\approx 5 E/(\hbar\omega)$. If this value of $N$ is reached, the effects from the finite cutoff are small
and the extrapolation is usually not required. If more bodies are considered, however,  it is computationally
more expensive to go to large values of $N$ and extrapolation becomes necessary.
In order to test the accuracy of our method, we have tried different extrapolations of our results
for the 2-body energies and compared to the exact solution.
The leading corrections from the finite cutoff $N$ scale as $1/\sqrt{N+3/2}$. A linear extrapolation
of our results in $1/\sqrt{N+3/2}$ works very well. If only the points with $N>80$ are taken into account,
linear and polynomial extrapolations give errors well below 1\%.

\subsection{Three and more bodies}

In order to extend this procedure to the three and higher-body sector, two additional tools are needed: 
the Talmi-Moshinsky-Transformation (TMT) \cite{brody} and  Wigner 6j symbols \cite{varshalovich}.
In the 3-body sector the finite model space is the linear hull of the tensor products 
$\phi_{n_1l_1m_1}\left(\vec{s}_1\right)\otimes\phi_{n_2l_2m_2}\left(\vec{s}_2\right)$ with the restriction 
$2\left(n_1+n_2\right)+l_1+l_2\leq N$. There are three 2-body contact interactions $V_i$ ($i=1,2,3$) and a  
3-body contact interaction $V^{\left(3\right)}_{\textrm{cont}}$. The contributions of these interactions are
most conveniently calculated in Jacobi coordinates $\vec{s}^{\left(i\right)}_1$ and $\vec{s}^{\left(i\right)}_2$,
$i=1,2,3$). There are three different sets which are labelled by the coordinate $i$ not appearing in 
the definition of $\vec{s}^{\left(i\right)}_1$.  The set
$\vec{s}^{\left(3\right)}_1$, $\vec{s}^{\left(3\right)}_2$ is given by
\begin{align*}
\vec{s}^{\left(3\right)}_1&=\frac{1}{\sqrt{2}}\left(\vec{x}_1-\vec{x}_2\right)\:,\\
\vec{s}^{\left(3\right)}_2&=\frac{1}{\sqrt{6}}\left(\vec{x}_1+\vec{x}_2-2\vec{x}_3\right)\:,
\end{align*}
and the other two combinations can be obtained by cyclically permuting the indices.
In order to transform the matrix elements between different sets of coordinates the TMT is used. The Talmi-Transformation 
transforms a set of coordinates ($\rho$, $\lambda$) to a set ($\rho'$, $\lambda'$) via
\begin{align*}
\begin{pmatrix}
	\vec{\rho'}\\
	\vec{\lambda'}\\				
\end{pmatrix}
=
\underbrace{
\begin{pmatrix}
\sqrt{\frac{1}{1+d}}&  -\sqrt{\frac{d}{1+d}}\\
\sqrt{\frac{d}{1+d}}&  \sqrt{\frac{1}{1+d}}\\
\end{pmatrix}}_{M}
\cdot
\begin{pmatrix}
	\vec{\rho}\\
	\vec{\lambda}\\				
\end{pmatrix}\:,
\end{align*}
where $d\geq0$ is a real number.
In our case, we need the transformation from one set of Jacobi coordinates to another set of Jacobi coordinates.
The Talmi transformation with $d=3$ transforms the set ($\vec{s}^{\left(3\right)}_1$, $\vec{s}^{\left(3\right)}_2$)
to the set ($-\vec{s}^{\left(1\right)}_1$, $-\vec{s}^{\left(1\right)}_2$) and the set 
($\vec{s}^{\left(3\right)}_1$, $-\vec{s}^{\left(3\right)}_2$) to the set ($-\vec{s}^{\left(2\right)}_1$, 
$\vec{s}^{\left(2\right)}_2$), respectively.
The corresponding expansion of the coupled oscillator functions 
depending on one set of Jacobi coordinates in oscillator functions 
depending on another set terminates after a finite number of terms, 
since the total oscillator energy, the total angular momentum, and the parity are conserved:
\begin{multline*}
	\left[\phi_{n_{\vec{\rho}}l_{\vec{\rho}}}\left(\vec{\rho}\right)\otimes\phi_{n_{\vec{\lambda}}l_{\vec{\lambda}}}
\left(\vec{\lambda}\right)\right]^L_{M_L}=
	\sum\limits_{n'_{\rho'}l'_{\rho'}n'_{\lambda'}l'_{\lambda'}}\left<n'_{\rho'}l'_{\rho'},
n'_{\lambda'}l'_{\lambda'};L|n_{\rho}l_{\rho},n_{\lambda}l_{\lambda}\right>_d\\
	\left[\phi^\beta_{n'_{\vec{\rho}\,'}l'_{\vec{\rho}\,'}}\left(\vec{\rho}\,'\right)\otimes\phi^\beta_{n'_{\vec{\lambda}'}
l'_{\vec{\lambda}'}}\left(\vec{\lambda}'\right)\right]^L_{M_L}\:.
\end{multline*}
The expansion coefficients $\left<n'_{\rho'}l'_{\rho'},
n'_{\lambda'}l'_{\lambda'};L|n_{\rho}l_{\rho},n_{\lambda}l_{\lambda}\right>_d$ are
the Brody-Moshinsky-Brackets. 

The coupling constants of the 2-body interactions are determined by Eq.~\eqref{2renorm} for a given 2-body energy. 
The coupling constant $v^{\left(3\right)}$ of $V^{\left(3\right)}_{\textrm{cont}}$ is determined so that a given 3-body 
energy $\epsilon^{\left(3\right)}$ is reproduced. Using the separability of $V^{\left(3\right)}_{\textrm{cont}}$ one finds for
the running coupling constant $v^{\left(3\right)}(N)$:
\begin{align}
\label{3renorm}
-\frac{\hbar\omega\pi^3}{v^{\left(3\right)}}=
\sum_{k}\frac{\bigl|\sum\limits_{n_{\vec{s}_1},n_{\vec{s}_2}}Z\left(k;n_{\vec{s}_1}0,n_{\vec{s}_2}0,0\right)
f_{n_{\vec{s}_1}}f_{n_{\vec{s}_2}}\bigr|^2}{D\left(k\right)-\epsilon^{\left(3\right)}}
\end{align}
with the eigenvector
\begin{align}
\left|k\right>=\sum\limits_{n'_{\vec{s}_1}l'_{\vec{s}_1},n'_{\vec{s}_2}l'_{\vec{s}_2},L}^{N}Z\left(k;n'_{\vec{s}_1}
l'_{\vec{s}_1},n'_{\vec{s}_2}l'_{\vec{s}_2},L\right)
\left|n'_{\vec{s}_1}l'_{\vec{s}_1},n'_{\vec{s}_2}l'_{\vec{s}_2},L\right>\:,
\end{align}
corresponding to the eigenvalue problem 
$D(k)\left|k\right>=\left(\frac{H^{\left(2\right)}_{\textrm{osc}}+V_1+V_2+V_3}{\hbar\omega}\right)\left|k\right>$.

Analogous to the 3-body sector, the model space for four bodies is the linear
hull of tensor products of HOF with
$2\left(n_1+n_2+n_3\right)+l_1+l_2+l_3\leq N$. In order to determine all
contributions of the interactions in one set of Jacobi-coordinates, the TMT
is used again. The transformation is defined for coupled wave functions
\begin{align*}
\left[\left[\phi_{n_{1}l_{1}}\left(\vec{s}_1\right)\otimes\phi_{n_{2}l_{2}}
\left(\vec{s}_2\right)\right]^{L_{12}}\otimes\phi_{n_{3}l_{3}}\left(\vec{s}_3
\right)\right]^L_M\:.
\end{align*}
For the TMT related to the mapping
$\{\vec{s}_2,\vec{s}_3\}\rightarrow\{\vec{s'}_2,\vec{s'}_3\}$, the wave
functions must at first be recoupled to
\begin{align*}
\left[\phi_{n_{1}l_{1}}\left(\vec{s}_1\right)\otimes\left[\phi_{n_{2}l_{2}}
\left(\vec{s}_2\right)\otimes\phi_{n_{3}l_{3}}\left(\vec{s}_3\right)\right]^{
L_{23}}\right]^{L'}_{M'}\:.
\end{align*}
The overlap between these wave functions is given by the 6j symbols
\cite{varshalovich}.
\begin{multline*}
\bigl<\left[\left[\phi_{n_{1}l_{1}}\otimes\phi_{n_{2}l_{2}}\right]^{L_{12}}
\otimes\phi_{n_{3}l_{3}}\right]^L_M\bigr|\left[\phi_{n_{1}l_{1}}\otimes\left[
\phi_{n_{2}l_{2}}\otimes\phi_{n_{3}l_{3}}\right]^{L_{23}}\right]^{L'}_{M'}\bigr>=
\\
\delta_{L,L'}\delta_{M,M'}\left(-1\right)^{l_1+l_2+l_3+L}\sqrt{\left(2L_{12}
+1\right)\left(2L_{23}+1\right)}
\begin{Bmatrix}
l_1 & l_2 & L_{12} \\ l_3 & L& L_{23}
\end{Bmatrix}\:.
\end{multline*}
This construction of the model space can be generalised to $n$ bodies in a straightforward way.

\section{Results for the 3-body sector}
\label{sec:3body}
\subsection{Unitary limit}

We are now in the position to apply this effective theory to the 3- and 4-body sector.
First, we focus on the 3-body sector in the unitary limit in order to test our method against the exact analytical solution \cite{werner}.
The corresponding ground state energy for the 2-body system 
is $\epsilon^{\left(2\right)}=1/2$ \cite{busch}. Since the three coupling constants $v^{\left(2\right)}_i$ 
are identical, the Hamiltonian is invariant under permutation of the bodies. It follows that the Hamilton matrix is 
block diagonal in different symmetries of the wave functions. Typical values of the cutoff $N$ reached in
our study for the 3-body problem are $N\approx 40...70$.

%%%%%%%%%%%%%%%%%%%%%%%%%%%%%%%%%%%%%%%%%%%%%%%%%%%%%%%%%%%%%%%%%%%%%%%%%%%%%%%%%%%%
\begin{figure}[t]
	\centering
		\includegraphics[width=0.5\linewidth, angle=270]{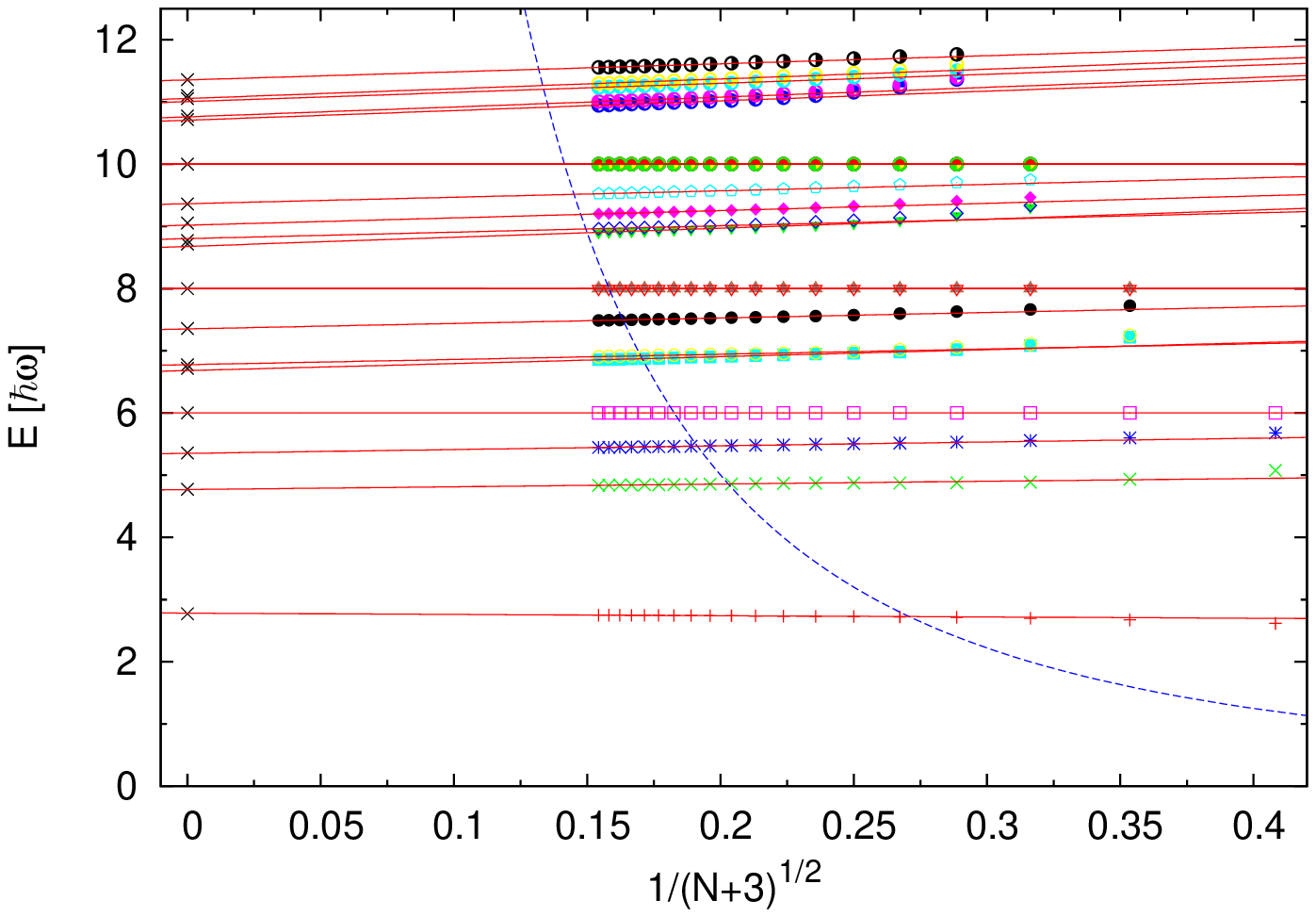}
	\caption{Energy spectrum for mixed antisymmetric states of three particles in a harmonic trap with $L^P=1^-$ as a function of $N$ 
          and exact analytical results from \cite{werner}. The solid lines indicate the linear extrapolation to $N=\infty$ 
           while the dashed line indicates the value of the cutoff $N$ at which $N\approx 5 E/(\hbar\omega)$.}
	\label{fig:3spectrumFL1-1}
\end{figure}
%%%%%%%%%%%%%%%%%%%%%%%%%%%%%%%%%%%%%%%%%%%%%%%%%%%%%%%%%%%%%%%%%%%%%%%%%%%%%%%%%%%%
For mixed antisymmetric states,
which are antisymmetric under exchange of the first two particles,
$V^{\left(3\right)}_{\textrm{cont}}$ does not contribute. As an example, 
Fig.~\ref{fig:3spectrumFL1-1} 
presents the spectrum of states with angular momentum and parity $L^P=1^-$ depending on the cutoff $N$ in the 
unitary limit. Again, the exact results are added at $N\rightarrow\infty$. 
The dashed line indicates the value of the cutoff $N$ at which $N\approx 5 E/(\hbar\omega)$. It is clear that for the higher excited
states an extrapolation in $N$ is required. The spectrum  can be extrapolated linearly in $1/\sqrt{N+3}$
as indicated by the solid lines.

\begin{figure}[t]
	\centering
		\includegraphics[width=0.5\linewidth, angle=270]{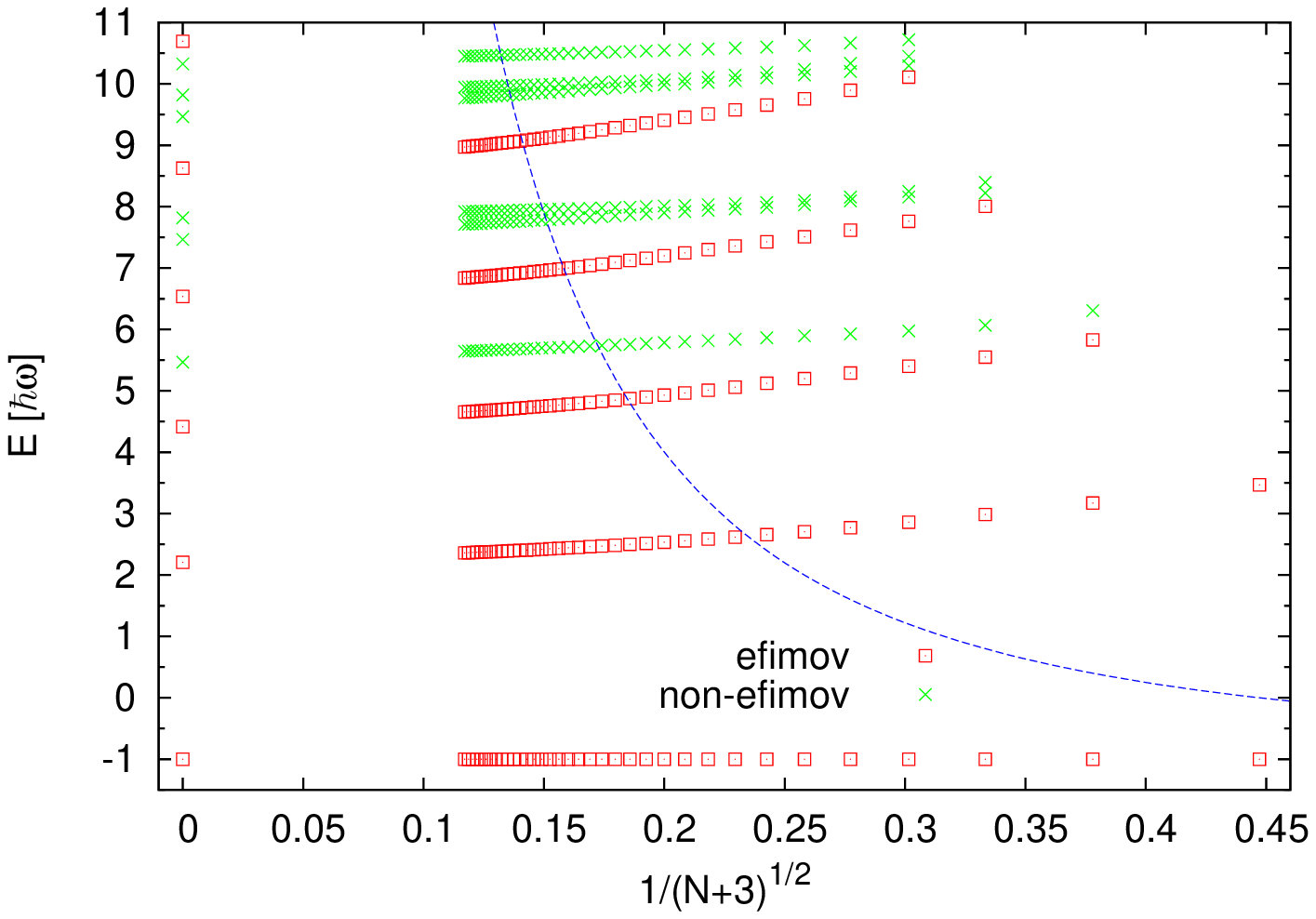}
	\caption{Energy spectrum for completely symmetric states of three particles in a harmonic trap with $L^P=0^+$ as a function of $N$ and exact 
                analytical results from \cite{werner}. Efimov-like states are indicated by squares, other states are
                indicated by crosses.
                The dashed line indicates the value of the cutoff $N$ at which $N\approx 5 E/(\hbar\omega)$.}
	\label{fig:3spectrumBL0-1}
\end{figure}

For completely symmetric wave functions, however, the 3-body interaction $V^{\left(3\right)}_{\textrm{cont}}$ does
contribute. Such wave functions can occur for systems of three identical bosons or three distinguishable particles
which both display the Efimov effect. In Fig.~\ref{fig:3spectrumBL0-1}, we show %(for $N\leq70$) 
the spectrum for positive parity and $L=0$ as a function of the cutoff $N$. 
As a typical example, the 3-body interaction 
was adjusted such that the 3-body ground state has energy $E_0^{(3)}\equiv \epsilon^{\left(3\right)}\hbar\omega=-\hbar\omega$. 
In principle, any value could be taken for $\epsilon^{(3)}$. However, 
the renormalization energies $\epsilon^{(3)}$ and $\epsilon^{(2)}$
should be well within the energy range given by the finite cutoff $N$ 
in order to avoid large errors from the finite cutoff.
As shown by Werner and Castin \cite{werner}, there are two different types of states. 
On the one hand, there are states independent of $V^{\left(3\right)}$ (crosses). On the other hand, there are states which 
depend on $V^{\left(3\right)}$ (squares) and are called Efimov-like. These Efimov-like states are the analog of Efimov
states in the trap. The exact results \cite{werner} are given at $N\rightarrow\infty$.
The dashed line indicates the value of the cutoff $N$ at which $N\approx 5 E/(\hbar\omega)$. Again, for the higher excited states
an extrapolation is required. For the non-Efimov-like states a linear extrapolation is appropriate.
The Efimov-like states, however, show a curvature. In this case, a quadratic term has to be included in the 
extrapolation. Typical extrapolation errors for Efimov-like states are of order 2-3\% and less than 1\% for the 
other states.
%%%%%%%%%%%%%%%%%%%%%%%%%%%%%%%%%%%%%%%%%%%%%%%%%%%%%%%%%%%%%%%%%%%%%%%%%%%%%%%%%%%%
%%%%%%%%%%%%%%%%%%%%%%%%%%%%%%%%%%%%%%%%%%%%%%%%%%%%%%%%%%%%%%%%%%%%%%%%%%%%%%%%%%%%

\subsection{Application to $^6$Li atoms}

\begin{figure}[t]
	\centering
		\includegraphics[width=0.5\linewidth, angle=270]{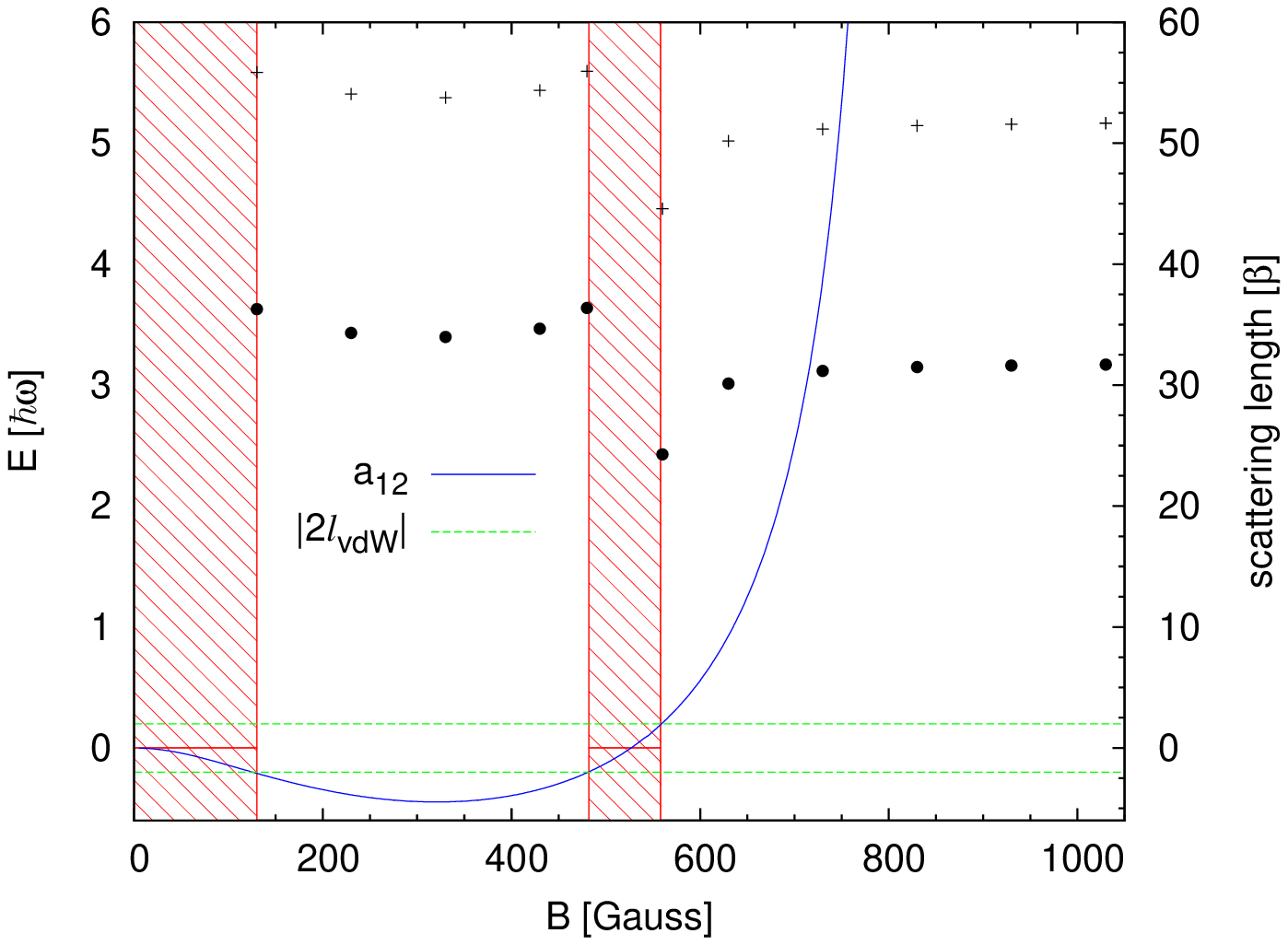}
	\caption{The energies of first two $0^+$ states in the 3-body system $|1\rangle\otimes |1\rangle\otimes |2\rangle$ 
                 indicated by circles and pluses (left scale) as function of the magnetic field. The solid and dashed
                 lines give the scattering length $a_{12}$ from \cite{julienne} and the van der Waals length 
                 $l_{\textrm{vdW}}$ (right scale).}
	\label{fig:3Li65L0ab}
\end{figure}

As an example application, we consider a system consisting of three fermionic $^6$Li atoms in a HOP with a low external 
magnetic field. Our calculations apply to experiments with a few fermions per site in optical lattices.
The scattering lengths for the three lowest hyperfine states labelled by their hyperfine 
quantum numbers $|f,m_f\rangle$, or by integers: $|1\rangle=|1/2,+1/2\rangle$, 
$|2\rangle=|1/2,-1/2\rangle$ and $|3\rangle=|3/2,-3/2\rangle$
have been calculated by Julienne \cite{julienne}.
The Feshbach resonances occur for magnetic fields near $834$ G, $811$ G and $690$ G. The corresponding renormalisation 
energies are determined by the relation between the ground state energy $\epsilon^{\left(2\right)}$ and the scattering 
length $a$ in the 2 particle sector (See Fig.\ref{fig:2ab}). The range of interactions for ultracold atoms is the 
van der Waals length $l_{\textrm{vdW}}$, which is approximately $62.5\,a_B$ for $^6$Li, where $a_B=0.529\,$\AA\ is the Bohr radius.
Thus the effective theory is only reliable for regions, where $a\simgt 2 l_{\textrm{vdW}}$.
A discussion of Efimov physics in $^6$Li atoms with the three spin states $|1\rangle$, $|2\rangle$, and $|3\rangle$ 
without trap was given in Refs.~\cite{Braaten:2009ey,nakajima2010,braaten2009,naidon2009,floerchinger2009}.
As an example, we calculate the 3-body spectrum for the system $|1\rangle\otimes |1\rangle\otimes |2\rangle$ 
which contains two like atoms in the trap. Because of the Pauli principle, there are no Efimov-like states 
and $V^{\left(3\right)}_{\textrm{cont}}$ does not contribute.  
The energy spectrum of two-component fermions in a trap as a function of the scattering length $a$ was previously studied in 
Refs.~\cite{chang,stecher:2007,blume:2007,Stetcu:2007ms,kestner:2007,Luu:2006xv,stecher:2008,Alhassid:2008,Stetcu:2010xq,Rotureau:2010uz,Drummond:2010}.
Figure~\ref{fig:3Li65L0ab} shows the extrapolated energies of the first two states with $L^P=0^+$ as a function of the magnetic 
field (circles and pluses). The oscillator length is taken as $\beta\approx65\,a_B$.
%%%%%%%%%%%%%%%%%%%%%%%%%%%%%%%%%%%%%%%%%%%%%%%%%%%%%%%%%%%%%%%%%%%%%%%%%%%%%%%%%%%%
%%%%%%%%%%%%%%%%%%%%%%%%%%%%%%%%%%%%%%%%%%%%%%%%%%%%%%%%%%%%%%%%%%%%%%%%%%%%%%%%%%%%
The solid and dashed lines give the scattering length $a_{12}$ from \cite{julienne} and the van der Waals length 
$l_{\textrm{vdW}}$ (right scale). The hatched areas indicate where the scattering length can not be considered
large and our theory is not applicable. Outside of the hatched areas, 
our calculation should be accurate up to corrections of order 
$l_{\textrm{vdW}}/a_{12}$ and $l_{\textrm{vdW}}/\beta$.
The magnetic field dependence of both states is very weak
despite the strong variation of the scattering length $a_{12}$.
The energies of the ground  and first excited state are of order $3.2\hbar\omega$ and $5.2\hbar\omega$,
respectively.

\section{Results for the 4-body sector}
\label{sec:4body}

We now turn to the 4-body system. We consider two types of 4-body systems both of which can display Efimov physics
and focus on the systematics of the bound state spectrum.
First, we consider two identical fermions with two other distinguishable particles. Because of the two 
identical fermions involved, this system is calculationally simpler than the 4-boson system which we calculate in
the second step. Typical values of the cutoff $N$ reached in
our study for the 4-body problem are $N\approx 20...30$.

\subsection{Two identical fermions and two distinguishable particles}

%%%%%%%%%%%%%%%%%%%%%%%%%%%%%%%%%%%%%%%%%%%%%%%%%%%%%%%%%%%%%%%%%%%%%%%%%%%%%%%%%%%%
\begin{figure}[t]
	\centering
			\includegraphics[width=0.5\linewidth,angle=270]{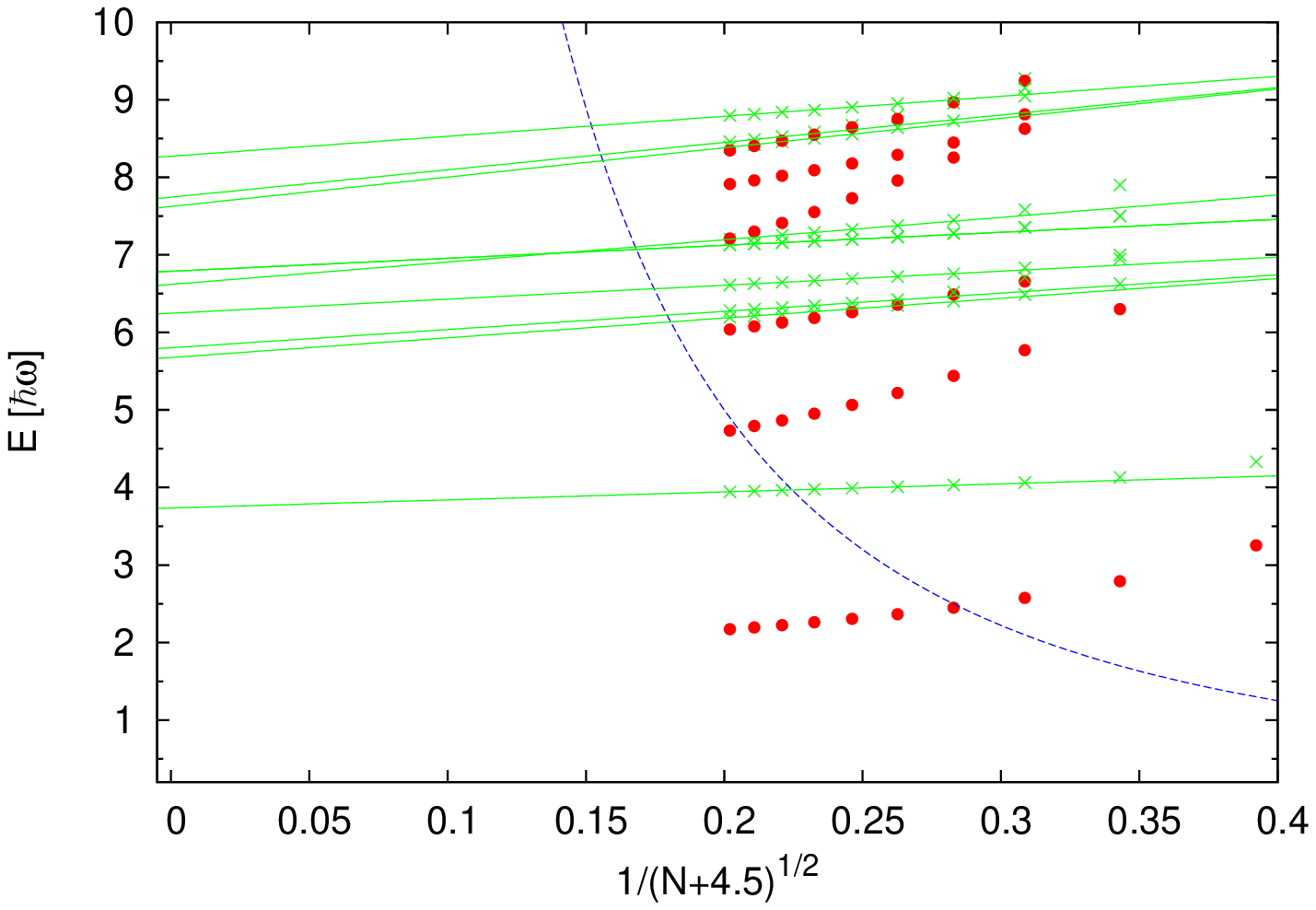}
	\caption{Energy spectrum for two identical fermions and two  distinguishable particles with $L^P=0^+$ in a harmonic trap 
                 as a function of $N$. The system is in the unitary limit and $\epsilon^{\left(3\right)}=-0.5$.
                 Efimov-like states are indicated by circles while universal states not 
sensitive to $\epsilon^{(3)}$ are given by crosses.
                     The solid lines indicate the linear extrapolation to $N=\infty$ 
                     while the dashed line marks the value of the cutoff $N$ at which $N\approx 5 E/(\hbar\omega)$.}
		\label{fig:4L0P00.5nlin}
\end{figure}
%%%%%%%%%%%%%%%%%%%%%%%%%%%%%%%%%%%%%%%%%%%%%%%%%%%%%%%%%%%%%%%%%%%%%%%%%%%%%%%%%%%%
We start with the case of two identical fermions with two other distinguishable particles. Thus, the wave function is 
antisymmetric under permutation of the identical fermions. Altogether, there are five 2-particle contact interactions 
and two 3-particle contact interactions. The contributions of some interactions are pairwise identical 
because two of the particles are identical. Analogous to the 3-body sector, the four coupling constants of the 2-particle 
interactions are renormalised by Eq.~\eqref{2renorm} and the two coupling constants of the 3-particle interactions by 
Eq.~\eqref{3renorm} for given ground state energies $\epsilon^{\left(2\right)}$ and $\epsilon^{\left(3\right)}$. 
We find both Efimov-like states which are sensitive to the value of $\epsilon^{\left(3\right)}$
and universal states which are independent of $\epsilon^{\left(3\right)}$.
At higher energies, we also find non-interacting eigenstates which are not sensitive
to both $\epsilon^{\left(2\right)}$ and $\epsilon^{\left(3\right)}$. 
As in the 2- and 3-body sector, the universal states 
%, which are independent of $\epsilon^{\left(3\right)}$ 
can be extrapolated linearly while the Efimov-like states require a quadratic extrapolation.
Here we require that no minimum exists for $1/N>0$ in order
to get stable results.
In Fig.~\ref{fig:4L0P00.5nlin}, we show our results for the spectrum states with $L^P=0^+$ in the unitary limit and for 
$\epsilon^{\left(3\right)}=-0.5$. The dashed line indicates the value of the cutoff $N$ at which $N\approx 5 E/(\hbar\omega)$.
For all but the lowest two states the condition $N\simgt 5 E/(\hbar\omega)$ is not satisfied for the values of $N$ reached in
this calculation. 
As a consequence, a stable extrapolation for $N\to\infty$ is required to obtain reliable results.
Since the exact results are not known, we estimate the uncertainty from the extrapolation conservatively as 
being equal to the energy shift from the last calculated value to the extrapolated value. This procedure should 
give an upper bound on the extrapolation uncertainty.

\begin{figure}[t]
	\centering
		\includegraphics[width=0.5\linewidth,angle=270]{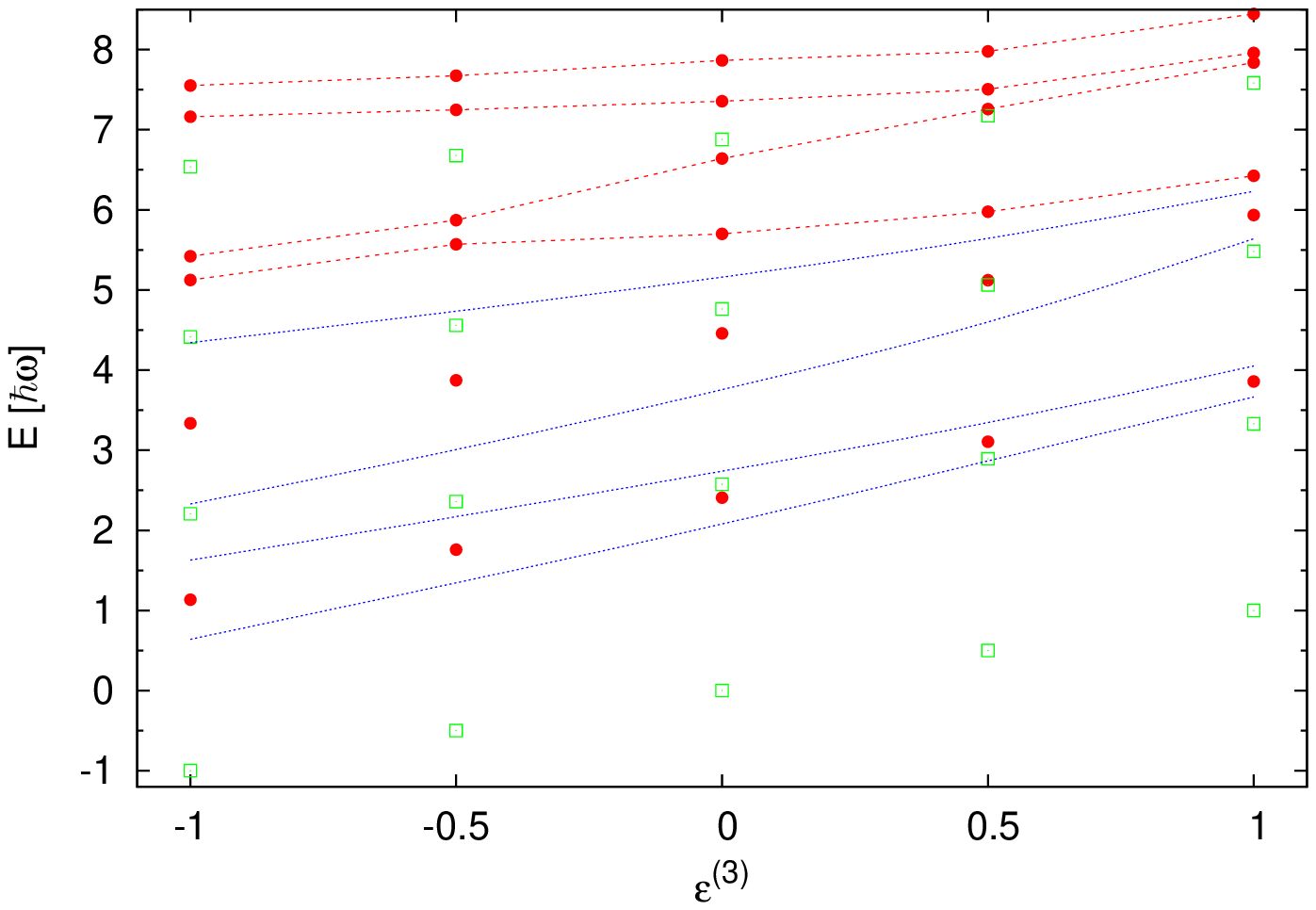}
	\caption{Extrapolated spectra for Efimov-like $0^+$ states in the 4-body sector of two identical fermions with two 
other distinguishable particles (circles) and in the 3-body sector (squares) for various $\epsilon^{\left(3\right)}$ in the unitary limit.
The dotted lines give our upper bound on the extrapolation error for the two lowest states. The dashed lines are guides to the eye. 
}
		\label{fig:four}
\end{figure}

In order to understand the systematics of the trapped spectra, we investigate the 4-body spectrum as a function of the 
input parameters $\epsilon^{\left(2\right)}$ and $\epsilon^{\left(3\right)}$ analogous to the Efimov plot
in free space \cite{Efimov-70}.
In Fig.~\ref{fig:four}, the extrapolated spectra for two identical fermions with two other distinguishable particles
in the unitary limit
are shown for various ground state energies $\epsilon^{\left(3\right)}$ by the circles. 
Additionally, the exact results for the Efimov-like states in the 3-body sector are shown (squares). 
Both the 3- and the 4-body states depend on $\epsilon^{\left(3\right)}$ approximately linearly in this region but the
slope of the 4-body states is typically larger. 
%%%%%%%%%%%%%%%%%%%%%%%%%%%%%%%%%%%%%%%%%%%%%%%%%%%%%%%%%%%%%%%%%%%%%%%%%%%%%%%%%%%%
%%%%%%%%%%%%%%%%%%%%%%%%%%%%%%%%%%%%%%%%%%%%%%%%%%%%%%%%%%%%%%%%%%%%%%%%%%%%%%%%%%%%
The dotted lines give our upper bound on the extrapolation error for the two lowest 4-body states. For the higher states
the extrapolation error is similar. The higher excited states are connected by dashed lines to guide the eye.
Note that the 4-body ground state is above the 3-body ground state in this case.
Around $\epsilon^{\left(3\right)}\approx 0.25$, the trajectories of lowest two and of the 4th 4-body state cross the next higher 3-body  
state. In free space, the corresponding 4-body state would become unstable to decay into the trimer and another particle, but in the 
trap this behavior can in principle be observed experimentally.
The other 4-body states are different in nature and move parallel with the nearest 3-body state and never cross.

\subsection{Four identical bosons}

\begin{figure}[t]
	\centering
		\includegraphics[width=0.5\linewidth,angle=270]{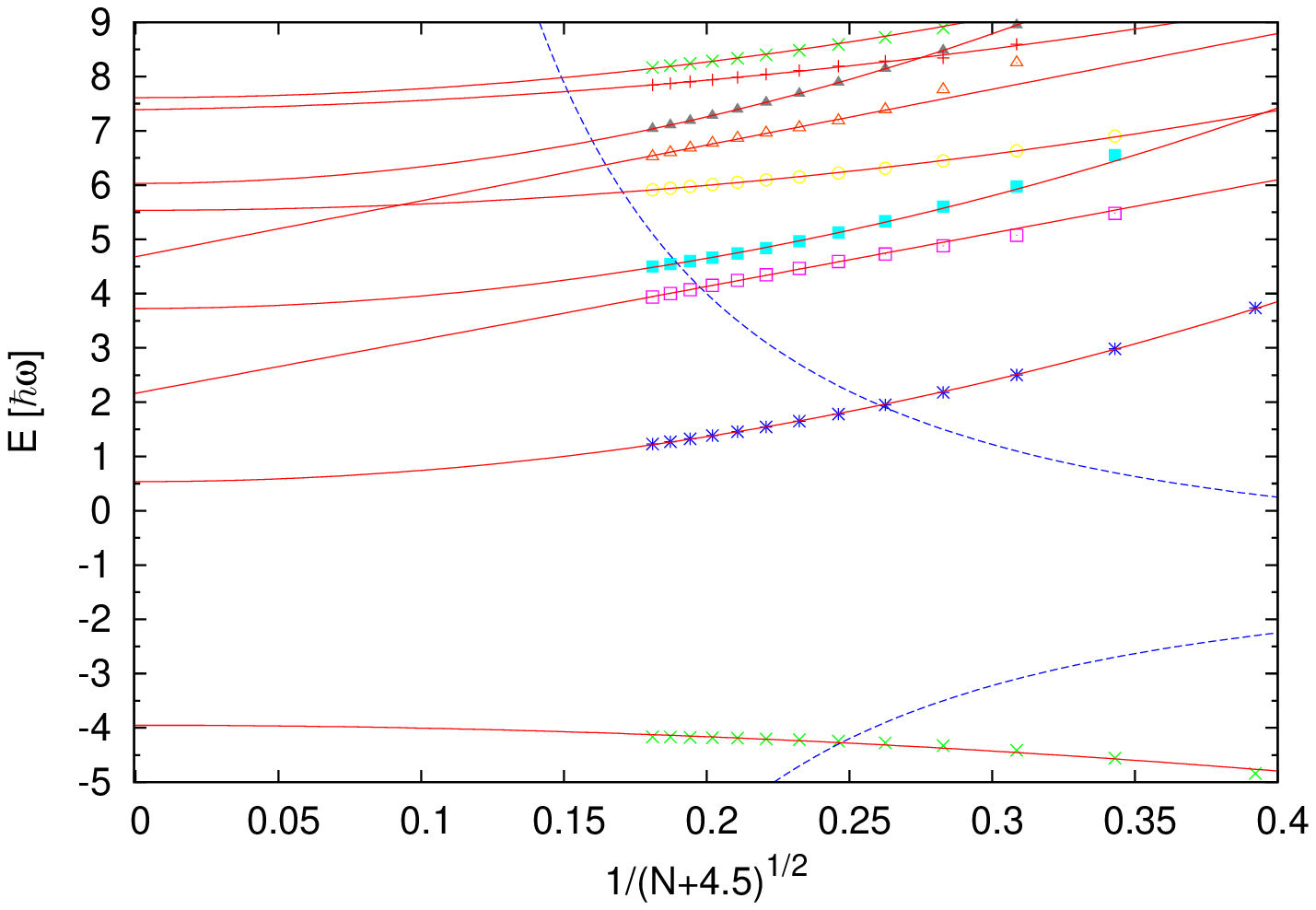}
	\caption{Spectrum for symmetric $0^+$ states of four identical bosons in the unitary limit for $\epsilon^{\left(3\right)}=-1$
          as a function of $N$. The solid lines indicate the extrapolation to $N=\infty$ 
           while the dashed line indicates the value of the cutoff $N$ at which $N\approx 5 E/(\hbar\omega)$.}
		\label{fig:4L0P0sym-1}
\end{figure}

We now turn to the system of four identical bosons with $L^P=0^+$. This system is of high experimental 
interest and the behavior of the states in free space is well known   \cite{Hammer:2006ct,Stecher:2008}.
Figure \ref{fig:4L0P0sym-1} shows the calculated spectrum in the unitary limit as a function of $N$ for 
$\epsilon^{\left(3\right)}=-1$. 
%%%%%%%%%%%%%%%%%%%%%%%%%%%%%%%%%%%%%%%%%%%%%%%%%%%%%%%%%%%%%%%%%%%%%%%%%%%%%%%%%%%%
%%%%%%%%%%%%%%%%%%%%%%%%%%%%%%%%%%%%%%%%%%%%%%%%%%%%%%%%%%%%%%%%%%%%%%%%%%%%%%%%%%%%
The dashed line indicates the value of the cutoff $N$ at which $N\approx 5 E/(\hbar\omega)$. 
For most of the states, the condition $N\simgt 5 E/(\hbar\omega)$ is not
satisfied and we therefore have to rely on the extrapolation.
Again, we find both Efimov-like states which are sensitive to the value of 
$\epsilon^{\left(3\right)}$ and universal states which are independent of $\epsilon^{\left(3\right)}$.
At higher energies, we also find non-interacting eigenstates which are insensitive
to both $\epsilon^{\left(2\right)}$ and $\epsilon^{\left(3\right)}$.
In the following, we focus on the Efimov-like states. 
They are extrapolated with a quadratic polynomial. Some of
the extrapolations are shown by the solid lines in Fig.~\ref{fig:4L0P0sym-1}. Again, 
we estimate the uncertainty from the extrapolation conservatively as 
being equal to the energy shift from the last calculated value to the extrapolated value.

\begin{figure}[t]
	\centering
	\includegraphics[width=0.5\linewidth,angle=270]{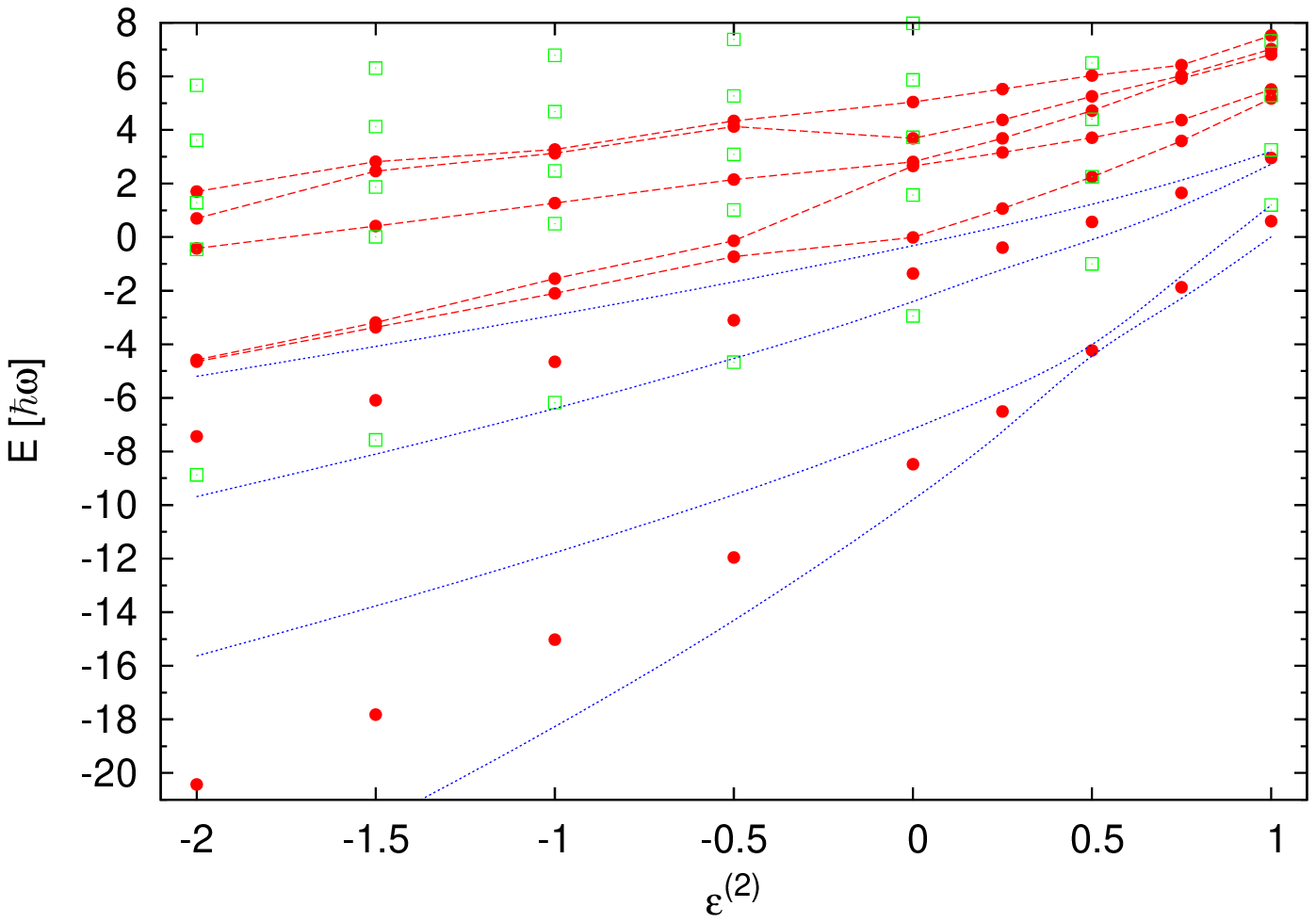}
	\caption{Extrapolated spectra of the symmetric 4-body states $0^+$ for various $\epsilon^{\left(2\right)}$ 
(circles) and 3-body Efimov-like states (squares). The 3-body interaction is fixed such that  $\epsilon^{\left(3\right)}=-1$
 in the unitary limit. The dotted lines give our upper bound on the extrapolation error for the two lowest states. 
The dashed lines are guides to the eye.}
		\label{fig:spectrumv123}
\end{figure}

We are now in the position to study the structure of the 3- and 4-body spectra for the 
Efimov-like states.
In the original Efimov plot, the 3-body spectrum is studied for fixed 3-body interaction $V^{\left(3\right)}$ while
the 2-body energy is varied \cite{Braaten:2004rn,Efimov-70}. Since there is no 4-body interaction
at leading order, this plot can be extended to the 4-body system and 
has been studied extensively in free space  \cite{Hammer:2006ct,Stecher:2008}. We will
compare our spectra with the free space results.
In Fig.~\ref{fig:spectrumv123}, the extrapolated spectra of the symmetric $0^+$ 4-body states for various 
$\epsilon^{\left(2\right)}$ are shown by the circles. The 3-body interaction is fixed by the requirement, that the 3-body ground 
state lies at $\epsilon^{\left(3\right)}=-1$ in the unitary limit. Additionally, the 3-body Efimov-like states 
are shown as squares.
%%%%%%%%%%%%%%%%%%%%%%%%%%%%%%%%%%%%%%%%%%%%%%%%%%%%%%%%%%%%%%%%%%%%%%%%%%%%%%%%%%%%
%%%%%%%%%%%%%%%%%%%%%%%%%%%%%%%%%%%%%%%%%%%%%%%%%%%%%%%%%%%%%%%%%%%%%%%%%%%%%%%%%%%%
As before, the dotted lines give our upper bound on the extrapolation error for the two lowest 4-body states. 
The higher excited states are connected by dashed lines to guide the eye.
Their extrapolation error is similar but not shown explicitly.
The harmonic confinement has a strong effect on the spectrum. Compared to free space, it is no longer true that two 4-body 
states are attached to each trimer state. Moreover, the levels appear to interact strongly. There are various avoided crossings of
4-body states, e.g. between the 4th and 5th state around $\epsilon^{\left(2\right)}\approx 0$
and possibly also between the second and third state. These avoided crossings 
could be studied experimentally by varying  $\epsilon^{\left(2\right)}$ using Feshbach resonances.

The dependence on $\epsilon^{\left(2\right)}$ can be translated into a dependence on the scattering length $a$ using
Fig.~\ref{fig:2ab}. For $\epsilon^{\left(2\right)}$ between minus two and zero, the scattering length is
essentially zero. When  $\epsilon^{\left(2\right)}$ is varied from zero to one, however, the scattering length grows to
become infinite at $\epsilon^{\left(2\right)}=1/2$, jumps to minus infinity and approaches a negative value close 
to zero at $\epsilon^{\left(2\right)}= 1$. This is the most interesting region from the point of universality
and corresponds to the usual Efimov plot in free space. In this region, the scattering length is much larger
than all other length scales and our effective theory is expected to describe systems of real atoms 
with van der Waals interactions.
The discrete scale invariance of the 3- and 4-body spectra in free space has disappeared in Fig.~\ref{fig:spectrumv123}. 
It would be interesting to approach the free space limit by making the trap wider and wider
but keeping the scattering length $a$ and the 3-body interaction fixed.
This corresponds to looking at the same physical system in a wider trap in order 
to see how the discrete scaling symmetry is restored. In the theoretical calculation, taking this limit is computationally
very expensive since the absolute value of the energy cutoff for fixed $N$ vanishes as $\beta\to\infty$. 
Cold atom experiments could serve as a quantum simulator to study this question.

\begin{figure}[t]
	\centering
	\includegraphics[width=0.5\linewidth,angle=270]{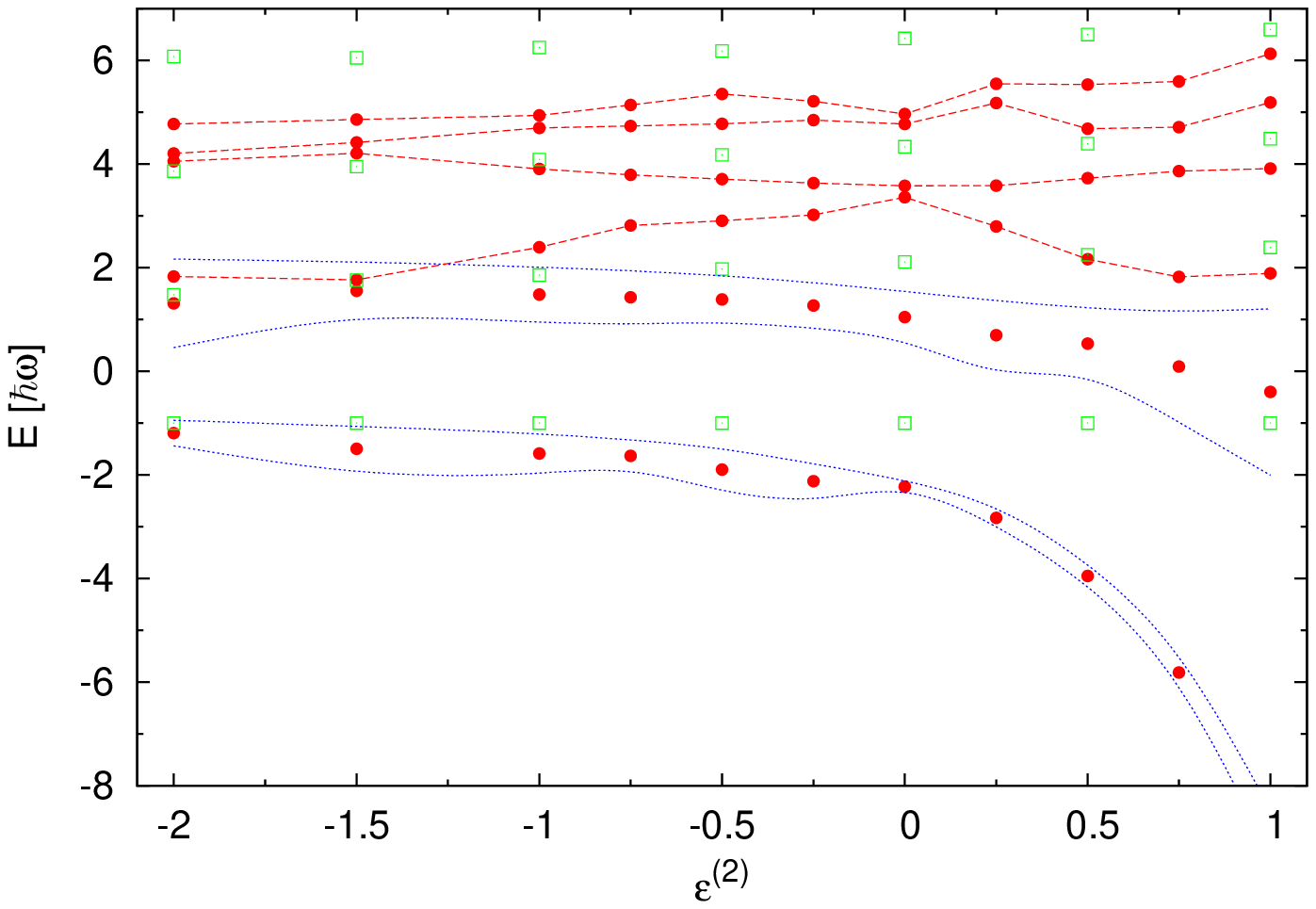}
	\caption{Extrapolated spectra of the symmetric 4-body (circles) and 3-body Efimov-like (squares) states $0^+$ 
for various $\epsilon^{\left(2\right)}$. For each $\epsilon^{\left(2\right)}$ the 3-body interaction is 
renormalised with the requirement, that the 3-body ground state lies at $\epsilon^{\left(3\right)}=-1$.
The dotted lines give our upper bound on the extrapolation error for the two lowest states. The dashed lines are guides to the eye.}
		\label{fig:spectrumsym}
\end{figure}

In Fig.~\ref{fig:spectrumsym}, we show a different variant of this plot.
The extrapolated spectra for 4-body $0^+$ states are 
given for various 2-body energies $\epsilon^{\left(2\right)}$ (circles). The 3-body interaction is renormalised with the 
requirement, that the 3-body ground state energy lies at $\epsilon^{\left(3\right)}=-1$ for each chosen 
$\epsilon^{\left(2\right)}$. Further, the Efimov-like 3-body states are added as squares.     
%%%%%%%%%%%%%%%%%%%%%%%%%%%%%%%%%%%%%%%%%%%%%%%%%%%%%%%%%%%%%%%%%%%%%%%%%%%%%%%%%%%%
%%%%%%%%%%%%%%%%%%%%%%%%%%%%%%%%%%%%%%%%%%%%%%%%%%%%%%%%%%%%%%%%%%%%%%%%%%%%%%%%%%%%
The extrapolation error estimates for the lowest two states are again given by the dotted lines. 
There are also avoided crossings for the higher states near $\epsilon^{\left(2\right)}\approx 0$.
In general, the spectrum is very different since the 3-body ground state is forced to be at $\epsilon^{\left(3\right)}=-1$. 
Experimentally, this situation is currently not accessible. It would require a second \lq\lq 3-body'' Feshbach resonance
that can be used to keep the 3-body ground state energy constant.

\section{Summary and Conclusions}
\label{sec:sum}

In this work, we have studied the universal properties of few-body
systems with large scattering length in a harmonic trap. We have 
used an effective theory for short-range interactions and identified
the cutoff on the harmonic oscillator basis functions with the 
cutoff of the effective theory \cite{Stetcu:2006ey}. The effective theory
is then formulated directly in the model space and the
low-energy constants are fixed by matching to few-body observables in the trap.
The few-body states in the trap are then obtained by diagonalization
of the finite Hamiltonian matrix. After renormalization, the spectrum 
is independent of the cutoff up to small corrections. If the cutoff 
can not be taken sufficiently large, it is possible to extrapolate 
to larger values. 

Our work has focused on systems where the Efimov effect occurs. 
In this case, a 3-body interaction already enters at leading order
and the corresponding low-energy constant has to be fixed by a 3-body
observable. We have checked our numerical method against exact 
analytical results for the 3-body system in the unitary limit and found
good agreement. Using the magnetic field dependence of the $^6$Li
scattering lengths calculated by Julienne \cite{julienne}, we have 
predicted the lowest two states in a two-state mixture as an example.

The main part of the work focused on the 4-body system in a harmonic
trap. We considered two types of 4-body systems both of which can display 
Efimov physics. First, we considered two identical fermions with two 
other distinguishable particles. 
Second, we calculated the spectrum
for four identical bosons. 
In both cases, we found both Efimov-like states which are sensitive to the value 
of $\epsilon^{\left(3\right)}$
and universal states which are independent of $\epsilon^{\left(3\right)}$.
At higher energies, there are also non-interacting eigenstates which are 
insensitive to both $\epsilon^{\left(2\right)}$ and $\epsilon^{\left(3\right)}$.
We found a number of interesting features in the spectra,
such as avoided crossings of 4-body states and 4-body states crossing
3-body states if the 2- or 3-body energies are increased. It would be very
interesting to observe these features experimentally. 
The trap introduces the oscillator length $\beta$ as a new scale and 
breaks the discrete scaling symmetry from free space.
In the calculated spectra no remnants of the discrete scaling symmetry 
were observed.

The extrapolation error of the calculated energies has been estimated conservatively 
from the difference of the last calculated value and the extrapolated value of the 
energy. These errors could be improved by going to larger values of $N$.\footnote{The 
present calculations were carried out on a laptop computer.}
However, it would also be desirable to understand the leading corrections from
the finite model space in more detail. Similar studies in free space have been carried 
out \cite{Ji:2010su}.

In general, the trap has a strong influence on the overall structure of the 
spectrum. In particular,
the discrete scaling symmetry and the connection of two 4-body states
to each trimer state known from free space could not be observed.
In the future it would be interesting to understand this modification 
in more detail by varying the size of the trap. Moreover, the extension 
to more bodies would allow to study the transition from few- to many-body
systems. While this is in principle straightforward, the computational effort 
grows substantially.

%\bibliography{literatur}
%\end{document}

% \appendix

% \section{}
% \label{}

% The Acknowledgements are also a un-numbered section
\section*{Acknowledgements}
% Acknowledgements text here
We thank D.R.~Phillips and U. van Kolck for discussions.
HWH was supported in part by the %Bundesministerium f\"ur Bildung und Forschung, 
BMBF under Contract No.~06BN9006.
He acknowledges the INT program ``Simulations and Symmetries: Cold Atoms, QCD, and
Few-Hadron Systems'', during which part of this work was carried out.

\end{document}